\documentclass[12pt]{CECS}
\usepackage{amsmath,amssymb,epsf,epsfig}

\title{Junction conditions in General Relativity with spin sources}
\author{Alex Giacomini, Ricardo Troncoso and Steven Willison\footnote{
emails: giacomini, steve, ratron at cecs.cl}
$ \medskip \medskip $ \\
{\small \emph{Centro de Estudios Cient\'{\i}ficos (CECS), Casilla
1469, Valdivia, Chile}.}}

\preprint{{\tiny CECS-PHY-06/05} }

\abstract{ The junction conditions for General Relativity in the
presence of domain walls with intrinsic spin are derived in three
and higher dimensions. A stress tensor and a spin current can be
defined just by requiring the existence of a well defined volume
element instead of an induced metric, so as to allow for generic
torsion sources. In general, when the torsion is localized on the
domain wall, it is necessary to relax the continuity of the
tangential components of the vielbein.

In fact it is found that the spin current is proportional to the
jump in the vielbein and the stress-energy tensor is proportional
to the jump in the spin connection. The consistency of the
junction conditions implies a constraint between the direction of
flow of energy and the orientation of the spin.

As an application, we derive the circularly symmetric solutions
for both the rotating string with tension and the spinning dust
string in three dimensions. The rotating string with tension
generates a rotating truncated cone outside and a flat space-time
with inevitable frame dragging inside. In the case of a string
made of spinning dust, in opposition to the previous case no frame
dragging is present inside, so that in this sense, the dragging
effect can be \textquotedblleft shielded" by considering spinning
instead of rotating sources. Both solutions are consistently
lifted as cylinders in the four-dimensional case.

}

\begin{document}

\section{Introduction}

The junction conditions for General Relativity have been studied first in a
systematic way in Ref. \cite{israel}, where one of the assumptions was that
the induced metric is well defined on the domain wall. In the first order
formalism, where the vielbein and the spin connection are varied
independently, this condition can be expressed, in a suitable frame, by the
fact that the vielbein should have no discontinuity tangential to the domain
wall.

It has been conjectured \cite{marolf} that, for reasonable matter
obeying certain energy conditions, singular shells should produce
a continuous metric. We should emphasise that this conjecture is
in the context of General Relativity without torsion.

However, it is shown here that the requirement of a continuous
metric is no longer consistent in general with torsion sources,
which are generated by spin currents along the domain wall. Thus,
the problem of finding the junction conditions for gravity in such
situations must be analyzed from scratch.

We derive the junction conditions for General Relativity in the
presence of domain walls with intrinsic spin in three and higher
dimensions. The new set of junction conditions reduce to Israel's
ones in the absence of torsion sources. We will show that a stress
tensor and a spin current can be defined by requiring the
existence of a well defined volume element instead of an induced
metric. This allows one to deal with situations where the torsion
is localized on the domain wall, where it is necessary to relax
the continuity of the tangential components of the vielbein. We
consider distributions of spin currents and stress-energy tensors
that can be obtained as the limit of a smooth distribution.
Consequently, the junction conditions consist of dynamical
equations relating the stress-energy tensor with the jump in the
spin connection, and the spin current with the jump in the
vielbein, as well as on certain constraints for the jump in the
geometry.

In order to see how the junction conditions work in a simple setup, we first
consider $2+1$ gravity without cosmological constant, where space-time is
flat outside the sources. The junction condition approach is then especially
suitable, being simply a matter of piecing together two flat manifolds.
Following this approach it is simple to obtain the circularly symmetric
solutions for both the rotating string with tension and the spinning dust
string in three dimensions.

Static closed string sources without torsion in $2+1$ gravity have
been studied in Ref. \cite{jackiwstring} solving the Einstein
field equations with a distributional source. Here, instead of
solving directly the Einstein field equations, we will extend the
static solution in two ways by using the new junction conditions.

We first discuss the extension to rotating case \cite{Clement},
where it is found that the metric outside corresponds to a
rotating truncated cone, while the inner metric describes flat
space-time with inevitable frame dragging. The matching conditions
impose for the rotating string a stress energy tensor of a fluid
with nonzero pressure. This non-static case therefore circumvents
the situation described in \cite{jackiwstring} where a closed
string with tension can only generate a cylinder space-time as
solution.

Next we discuss the case of a closed static string made of
spinning dust, which has a homogeneous torsion distribution on it.
As torsion is related to the spin density \cite{jackiwlectures} we
will refer to this as the \textquotedblleft spinning string". In
the case of a string made of spinning dust, the metric outside is
the same as for the rotating case, but in this case no frame
dragging is present inside.

It is worth pointing out that the presence of torsion concentrated on the
string forces a tangential discontinuity in the vielbein, but the induced
volume element is well-defined. Here we have a concrete example of a
physical situation where an extended gravitating object can be treated even
though there is no a well defined well defined induced metric.

From this concrete example, one concludes that having a spinning
string as consistent solution of the field equation is generically
incompatible with having a continuous metric. This is apparent
from the fact that the discontinuity in the vielbein cannot be
avoided by absorbing the singularity of the torsion in a $\delta
$-distribution in the spin connection, since in this case, the
Einstein tensor would acquire a delta function squared, and thus
the solution would be meaningless in the distributional sense.
This example then shows the need for generalizing the standard
junction conditions.

As an example of the junction conditions in higher dimensions, we show that
the two solutions can be consistently lifted to the rotating and spinning
cylinder in four dimensions in a similar way a point particle solution in $%
2+1$ is lifted to a cosmic string \cite{cosmicstring1}, \cite{cosmicstring2}%
. The lifting to higher dimensions is trivial.

The plan of the paper is as follows: In Section II we consider the
junction conditions in 2+1 dimensions, and then apply them to
obtain the rotating and spinning strings. Section III is devoted
to the four- and higher-dimensional case, where the junction
conditions are used to obtain the lifting of the solutions
previously discussed. Section IV contains the conclusions.

\section{Junction conditions for torsion sources in three dimensions}

In the three-dimensional case the junction conditions can be derived in a
straightforward way, and it is instructive to discuss it first in order to
gain insights about how to proceed in higher dimensions.

\subsection{2+1 General Relativity in first order formalism}

The Lagrangian for General Relativity in first order formalism in 2+1
dimensions is
\begin{equation*}
\mathcal{L}=\Omega ^{ab}\wedge e^{c}\epsilon _{abc}+\mathcal{L}_{\text{matter%
}}.
\end{equation*}%
where $e^{a}=e_{\mu }^{a}dx^{\mu }$ is the vielbein one-form, $\omega ^{ab}$
is the spin connection and $\Omega ^{ab}=d\omega ^{ab}+\omega _{\
c}^{a}\wedge \omega ^{cb}$ denotes the curvature 2-form.

The Euler-Lagrange variation with respect to the vielbein $e_{\mu }^{a}$
gives the Einstein equations:
\begin{gather}
\Omega ^{ab}\wedge dx^{\mu }\epsilon _{abc}=-2\mathcal{T}_{c}^{\mu }\text{Vol%
}\ ,  \label{Einstein} \\
\text{Vol}:=\frac{1}{3!}e^{a}\wedge e^{b}\wedge e^{c}\epsilon _{abc}=\sqrt{-g%
}d^{3}x  \notag
\end{gather}%
Here $\mathcal{T}_{c}^{\mu }$ is the stress-energy tensor. The more familiar
form is $\mathcal{T}_{\nu }^{\mu }=e_{\nu }^{c}\mathcal{T}_{c}^{\mu }$.

The Euler-Lagrange equation with respect to the spin connection $\omega
_{\mu }^{ab}$ gives%
\begin{equation}
dx^{\mu }\wedge T^{c}\epsilon _{abc}=\mathcal{S}_{ab}^{\mu }\text{Vol}
\label{Torsion}
\end{equation}%
where $T^{a}=De^{a}:=de^{a}+\omega _{\ b}^{a}\wedge e^{b}$ is the torsion,
and $\mathcal{S}_{ab}^{\mu }$ is the spin current. In any region of
space-time where there is no matter, these field equations imply that both
the curvature and torsion vanish. Note that we chose units where $8\pi G=1$.
\newline

\subsection{Junction conditions for a string}

We will consider a string which is an object of $1$ spatial dimension
supporting a singular distribution of matter possibly with spin current. The
$1+1$-dimensional worldsheet of the string will be denoted by $\Sigma $,
which divides the spacetime into two disconnected pieces $M_{+}$ and $M_{-}$%
, and it is assumed not to be a null surface. Consider a region, $O$, of
space-time which contains $\Sigma $ and is of arbitrarily small width in the
direction normal to $\Sigma $. The stress-energy distribution $(\mathcal{T}%
_{\Sigma })_{a}^{\mu }$ with support on the surface $\Sigma $ is defined by:
\begin{equation}
\lim_{O\rightarrow \Sigma }\int_{O}\mathcal{T}_{a}^{\mu }\text{Vol}%
=\int_{\Sigma }(\mathcal{T}_{\Sigma })_{a}^{\mu }\text{Vol}_{\Sigma }\ ,
\label{DistributionStress}
\end{equation}%
where $\text{Vol}_{\Sigma }$ stands for the induced volume element on $%
\Sigma $. Alternatively we can say $\mathcal{T}_{a}^{\mu }=(\mathcal{T}%
_{\Sigma })_{a}^{\mu }\delta (\Sigma )$ where $\delta (\Sigma )$ is the
Dirac delta function. Similarly, we define the spin current distribution to
be
\begin{equation}
\lim_{O\rightarrow \Sigma }\int_{O}\mathcal{S}_{ab}^{\mu }\text{Vol}%
=\int_{\Sigma }(\mathcal{S}_{\Sigma })_{ab}^{\mu }\text{Vol}_{\Sigma }.
\label{DistributionSpin}
\end{equation}

As matter is confined on $\Sigma $, the geometry is necessarily non-smooth.
As usual, the stress-energy of the string will generate a jump in the spin
connection. However, it can be seen that the concentration of torsion on the
string requires also a discontinuity in the vielbein at $\Sigma $. This is
further discussed in section \ref{nogo}. So, instead of requiring that the
geometry be continuous, we impose a weaker condition: we shall require only
that the vielbein and spin connection have a \emph{bounded} discontinuity at
$\Sigma $.

We integrate the l.h.s of the Einstein field equations
(\ref{Einstein}) over the region $O$. In the limit, only the
exterior derivatives of the spin connection contributes, giving
the following boundary term,
\begin{gather}
\lim_{O\rightarrow \Sigma }\int_{O}\Omega ^{ab}\wedge dx^{\mu }\epsilon
_{abc}=\lim_{O\rightarrow \Sigma }\int_{O}(d\omega ^{ab}+\cdots )\wedge
dx^{\mu }\epsilon _{abc}=\lim_{O\rightarrow \Sigma }\int_{O}d(\omega
^{ab}\wedge dx^{\mu }\epsilon _{abc})+(\cdots )  \notag \\
=\int_{\Sigma _{+}}\omega ^{ab}\wedge dx^{\mu }\epsilon _{abc}-\int_{\Sigma
_{-}}\omega ^{ab}\wedge dx^{\mu }\epsilon _{abc}
\label{Curvature_integration} \\
=\int_{\Sigma }\Delta \omega ^{ab}\wedge dx^{\mu }\epsilon _{abc}\ ,  \notag
\end{gather}%
where $\Delta \omega ^{ab}=\omega _{+}^{ab}-\omega _{-}^{ab}$ is the
discontinuity in the spin connection across $\Sigma $, and the dots $(\cdots
)$ represents those terms which vanish in the limit. Proceeding in the same
way with the torsion equation (\ref{Torsion}) one obtains the following
boundary term
\begin{gather}
\lim_{O\rightarrow \Sigma }\int_{O}dx^{\mu }\wedge T^{c}\epsilon
_{abc}=\lim_{O\rightarrow \Sigma }\int_{O}dx^{\mu }\wedge (de^{c}+\cdots
)\epsilon _{abc}=\lim_{O\rightarrow \Sigma }\int_{O}d(dx^{\mu }\wedge
e^{c}\epsilon _{abc})+(\cdots )  \notag \\
=\int_{\Sigma _{+}}dx^{\mu }\wedge e^{c}\epsilon _{abc}-\int_{\Sigma
_{-}}dx^{\mu }\wedge e^{c}\epsilon _{abc}  \label{Torsion_integration} \\
=\int_{\Sigma }dx^{\mu }\wedge \Delta e^{c}\epsilon _{abc}  \notag
\end{gather}%
where $\Delta e^{a}=e_{+}^{a}-e_{-}^{a}$ is the discontinuity in the
vielbein.

So, comparing (\ref{Curvature_integration}) and (\ref{Torsion_integration})
with (\ref{DistributionStress}) and (\ref{DistributionSpin}), we obtain the
junction conditions:

\begin{align}
i^{\ast }(\Delta \omega ^{ab}\wedge dx^{\mu }\epsilon _{abc})& =-2(\mathcal{T%
}_{\Sigma })_{c}^{\mu }\text{Vol}_{\Sigma }\ ,  \label{curvature_junction} \\
i^{\ast }(dx^{\mu }\wedge \Delta e^{c}\epsilon _{abc})& =(\mathcal{S}%
_{\Sigma })_{ab}^{\mu }\text{Vol}_{\Sigma }\ ,  \label{torsion_junction}
\end{align}%
where $i^{\ast }$ denotes the pull-back of differential forms to the surface
$\Sigma $.

According to Eq. (\ref{torsion_junction}), one can see that in the
absence of spin currents on $\Sigma $, the tangent components of
the vielbein are continuous, and thus, Eq
(\ref{curvature_junction}) reduces to the standard Israel Junction
conditions. Thus, we conclude that the presence of spin current on
$\Sigma $ in (\ref{torsion_junction}) necessarily forces the
discontinuity in the tangential components of the vielbein, i.e.
the induced metric on the string worldsheet $h_{\mu \nu }^{+}$
induced by the geometry in $M_{+}$ is different to $h_{\mu \nu
}^{-}$ induced by $M_{-}$.

Note that on the right hand side of the junction conditions, the
volume element of the string appears explicitly. This comes from
the definitions (\ref{DistributionStress}) and
(\ref{DistributionSpin}). In order for this to make sense, we
clearly need the intrinsic volume
element on $\Sigma $ to be single valued, i. e.,%
\begin{equation}
\text{Vol}_{\Sigma _{+}}=\text{Vol}_{\Sigma _{-}}\equiv
\text{Vol}_{\Sigma }\ . \label{volume_condition}
\end{equation}

In terms of the metric, equation \ref{volume_condition} means
that, although $h_{\mu \nu }^{+}\neq h_{\mu \nu }^{-}$, the
determinants are equal: $h_{+}=h_{-}$. In terms of the vielbein
this means that $i^{\ast }e_{+}^{a}\wedge e_{+}^{b}=i^{\ast
}e_{-}^{a}\wedge e_{-}^{b}$ even though $e_{+}^{a}\neq e_{-}^{a}$.
Therefore instead of requiring a well defined induced metric one
we can relax to the weaker condition of having a well defined
volume.

Upon a more careful treatment of the distributional field
equations, we will find in section \ref{higher_chapter} that extra
conditions are needed. In the case of 2+1 dimensions, there is one
extra condition $i^*\Delta e^{[a} \wedge \Delta e^{b]} = 0$.

Note that even in the absence of an induced metric on $\Sigma $, the
junction conditions (\ref{curvature_junction}), (\ref{torsion_junction}) can
be written in terms of purely anholonomic indices,
\begin{gather*}
i^{\ast }(\Delta \omega ^{ab}\wedge e^{d}\epsilon _{abc})=-2(\mathcal{T}%
_{\Sigma })_{c}^{d}\text{Vol}_{\Sigma }, \\
i^{\ast }(e^{d}\wedge \Delta e^{c}\epsilon _{abc})=(\mathcal{S}_{\Sigma
})_{ab}^{d}\text{Vol}_{\Sigma }.
\end{gather*}%
provided the following conditions are satisfied%
\begin{gather}
i^{\ast }(\Delta \omega ^{\lbrack ab}\wedge \Delta e^{d]})=0,  \label{extra1}
\\
i^{\ast }(\Delta e^{[d}\wedge \Delta e^{c]})=0.  \label{extra2}
\end{gather}%
This can be seen contracting Eqs. (\ref{curvature_junction}) and (\ref%
{torsion_junction}) with $e_{\mu }^{d}$, and requiring for the
left hand side of these equations to be independent wether we
contract with $e_{+}^{d}$ or $e_{-}^{d}$. Note that equation
(\ref{extra2}) is the same as the condition in the above
paragraph.

It is important to stress that the fundamental equations are the
junction conditions (\ref{curvature_junction}) and
(\ref{torsion_junction}), plus the conditions
(\ref{volume_condition}) and (\ref{extra2}). The other condition,
(\ref{extra1}) should be regarded as a weaker alternative to
having an induced vielbein on $\Sigma $ for the purposes of
changing to anholonomic indices. This extra condition does not
come from the field equations. We will see that this is not the
case in higher dimensions, where the condition (\ref{extra1}) is
strictly needed.

In the presence of a cosmological constant, the junction conditions are the
same since the cosmological constant term does not contain derivatives.

To summarize, the two junction conditions (\ref{curvature_junction}) and (%
\ref{torsion_junction}), together with (\ref{Einstein}) and (\ref{Torsion})
in the interior of $M_{+}$ and $M_{-}$, determine completely the solution
for the string.

\subsection{The static dust string}

\label{static}

The results obtained from the junction conditions agree with
solving the Einstein equation with delta function sources. This
can be seen explicitly for a static closed dust string
\cite{deserjackiw}. In this case the interior is a piece of
Minkowski space and that the exterior is locally Minkowski, with
the spatial section having the shape of a cone of deficit angle
$2\pi (1-B)$. A simple way to derive the mass of the string in
terms of $B$ is to write the metric for the whole space-time as
\begin{equation*}
ds^{2}=-dt^{2}+\left\{ 1-(1-B^{-2})\theta (r-r_{0})\right\}
dr^{2}+r^{2}d\phi ^{2}.
\end{equation*}%
Above, $\theta $ is the Heaviside distribution, which takes value $0$ for $%
r<r_{0}$ and $1$ for $r>r_{0}$. The location of the string will be $r=r_{0}$%
. Although in this form the metric appears discontinuous, this is an
artefact of the choice of coordinates\footnote{%
The coordinates outside the source can be chosen so that the
spatial
section of exterior metric is conformally flat%
\begin{equation*}
\left( \frac{r}{r_{0}}\right) ^{-{2}(1-B)}\left( dr^{2}+r^{2}d\phi
^{2}\right) \ ,
\end{equation*}%
and in this way the metric and the normal are continuous across
$\Sigma$.}. Indeed, the induced metric on the surface $r=r_{0}$ is
well defined:
\begin{equation*}
ds_{\Sigma }^{2}=-dt^{2}+r_{0}^{2}d\phi ^{2},
\end{equation*}%
which is the metric of the cylinder with radius $r_{0}$. We can choose the
vielbein to be:
\begin{align*}
e^{0}& =dt, \\
e^{1}& =\left\{ 1-(1-B^{-1})\theta (r-r_{0})\right\} dr, \\
e^{2}& =rd\phi .
\end{align*}%
Using the zero torsion condition, we find the non-vanishing component of the
spin connection:
\begin{equation}
\omega ^{12}=-\left\{ 1-(1-B^{-1})\theta (r-r_{0})\right\}
^{-1}d\phi =-\left\{ 1-(1-B)\theta (r-r_{0})\right\} d\phi .
\label{omega12}
\end{equation}%
The only non-trivial curvature component is:
\begin{equation*}
\Omega ^{12}=d\omega ^{12}=(1-B)\delta (r-r_{0})dr\wedge d\phi .
\end{equation*}%
Here $\delta $ is the Dirac delta distribution. Inserting this ansatz into
the field equations (\ref{Einstein}) we obtain%
\begin{equation*}
2(1-B)\delta (r-r_{0})dr\wedge d\phi \wedge dt=-2\mathcal{T}%
_{0}^{0}dt\wedge e^{1}\wedge e^{2}.
\end{equation*}%
The integration constant $B$ is related with the mass which can be
defined as the integral of $-\mathcal{T}_{0}^{0}e^{1}\wedge e^{2}$
over a spatial cross section
\begin{equation*}
M=\int drd\theta (1-B)\delta (r-r_{0})=2\pi (1-B)\ ,
\end{equation*}%
and so the mass (or $8\pi G$ times the mass, restoring Newton's
constant) is equal to the deficit angle.

As a check, we apply the junction conditions to the dust string to rederive
the above results. The discontinuity in the connection is $\Delta \omega
^{12}=(1-B)d\phi $. Putting this into the junction conditions (\ref%
{curvature_junction}) and (\ref{torsion_junction}), we get
\begin{equation*}
(\mathcal{T}_{\Sigma })_{0}^{0}=-(1-B)/r_{0}\ ,
\end{equation*}%
and the spin current vanishes. Integrating this around the strings
length, we get $M=2\pi (1-B)$ as expected. Note that the junction
conditions work even for a discontinuous vielbein. This
discontinuity contributes nothing to the torsion because it is
purely normal to the string i.e. $i^{\ast }\Delta e^{a}=0$.

\subsection{The rotating string with tension}
\label{rotating_section}

 Here we extend the previous result to the
rotating case with tension making
use of the junction conditions (\ref{curvature_junction}), (\ref%
{torsion_junction}). For the exterior region $M_{+}$, the metric
corresponds to the one of a rotating conical
spacetime\cite{deserjackiw}. This can be written in the familiar
form
\begin{equation}
ds_{+}^{2}=-\left( dt-\frac{J}{2}d\phi \right) ^{2}+d\tilde{r}^{2}+B^{2}%
\tilde{r}^{2}d\phi ^{2}\ ,  \label{Rotating-Cone-Metric}
\end{equation}%
In this section we will, by a rescaling $\tilde{r} \to
r(\tilde{r})$, write the metric in the following form
\begin{equation}
ds_{+}^{2}=-\left( dt-\frac{J}{2}d\phi \right) ^{2}+
(1+\chi^2)\left(\frac{r}{r_0} \right)^{2(B-1)}(d{r}^{2}+
{r}^{2}d\phi ^{2})\ , \label{Rotating-Cone-conformal}
\end{equation}%
where the constant $\chi$ is defined as
\begin{gather*}
\chi:= \frac{J}{2r_0}
\end{gather*}
and we choose the vielbein as
\begin{equation}
e_{+}^{0}=dt-\frac{J}{2}d\phi \ ,\ e_{+}^{1}=\sqrt{1+\chi^2}
\left(\frac{r}{r_0} \right)^{(B-1)}d{r}\ ,\ e_{+}^{2}=
\sqrt{1+\chi^2} \left(\frac{r}{r_0} \right)^{(B-1)}{r}d\phi \ .
\label{Rotating-Cone-Vielbein}
\end{equation}%
The string is located at ${r}={r}_{0}$ so that the induced metric
on its worldsheet is:
\begin{equation*}
-dt^{2}+J\ d\phi dt+r_{0}^{2}d\phi ^{2},
\end{equation*}
which ensures the absence of closed timelike curves in the outer
region.

The interior, $M_{-}$, is a region of Minkowski space,
\begin{equation*}
ds_{-}^{2}=-{dt^{\prime }}^{2}+dr^{2}+r^{2}{d\phi ^{\prime }}^{2}.
\end{equation*}%
Assuming that there is no spin current on the string, by virtue of (\ref%
{torsion_junction}) the induced metric is continuous. As a
consequence, the inner and outer frames and co-ordinates must be
related. Thus, the
matching of the coordinates gives%
\begin{equation*}
t^{\prime }=\sqrt{1+\chi ^{2}}\ t,\qquad \phi ^{\prime }=\phi +\frac{\chi }{%
r_{0}}\,t,
\end{equation*}%
where $\phi ^{\prime }$ has the same periodicity as $\phi $, and the
location of the string measured with respect to the inner coordinate $r$ is $%
r=r_{0}$.

The vielbein of the interior region can taken to be:
\begin{equation*}
e_{-}^{\hat{0}}=dt^{\prime },e_{-}^{\hat{1}}=dr,e_{-}^{\hat{2}}=rd\phi
^{\prime },
\end{equation*}%
so that the inner and outer induced vielbeins on the string worldsheet are
\begin{gather}
\theta _{-}^{\hat{0}}=\sqrt{1+\chi ^{2}}dt,  \label{induced minus} \\
\theta _{-}^{\hat{2}}=r_{0}d\phi +\chi dt.  \notag
\end{gather}%
and
\begin{gather}
\theta _{+}^{0}=dt-r_{0}\chi d\phi ,  \label{induced_plus} \\
\theta _{+}^{2}=r_{0}\sqrt{1+\chi ^{2}}d\phi ,  \notag
\end{gather}%
respectively. They are related by a Lorentz transformation\footnote{%
Here hatted indices correspond to the inner frame. This is useful because
the junction conditions are formulated in a single frame.}:
\begin{equation*}
\theta _{+}^{a}=\Lambda _{\ \hat{b}}^{a}\theta _{-}^{\hat{b}},\qquad \Lambda
_{\ \hat{b}}^{a}=%
\begin{pmatrix}
\sqrt{1+\chi ^{2}} & -\chi &  \\
-\chi & \sqrt{1+\chi ^{2}} &
\end{pmatrix}%
.
\end{equation*}

The spin connection in the interior is
\begin{equation*}
\omega _{-}^{\hat{1}\hat{2}}=-d\phi ^{\prime }
\end{equation*}%
and in the exterior region it is
\begin{equation*}
\omega _{+}^{12}=-Bd\phi .
\end{equation*}

Now we notice that $i^{\ast }\omega $ transforms like a tensor under the two
dimensional Lorentz transformations on the worldsheet \footnote{%
Since the intrinsic spin connection on the worldsheet vanishes on
both frames, for the above choice of vielbeins, $i^{\ast }\omega
_{\pm }^{ab}$ is the second fundamental form of the string
worldsheet with respect to the embedding into the exterior and
interior regions respectively. Thus, (\ref{same_as_Israel}) is the
same as the Israel junction condition. For a more general
situation, we would need to calculate $i^{\ast }(\omega
_{+}-\omega _{0})^{ab}$ and $i^{\ast }(\omega _{-}-\omega
_{0})^{\hat{a}\hat{b}}$, where $\omega _{0}$ is the connection
associated with the intrinsic geometry of the worldsheet.}.

The junction conditions give
\begin{equation}
i^{\ast }\left( \omega _{+}^{ab}\wedge \theta _{+}^{d}\epsilon _{abc}-\omega
_{-}^{\hat{a}\hat{b}}\wedge \theta _{-}^{\hat{d}}\epsilon _{\hat{a}\hat{b}%
\hat{c}}\Lambda _{\ \hat{d}}^{d}(\Lambda ^{-1})_{\ c}^{\hat{c}}\right) =-2%
\mathcal{T}_{c}^{d}\text{Vol}_{\Sigma }.\label{same_as_Israel}
\end{equation}

Using the fact that the volume element on the string worldsheet is $%
r_{0}dt^{\prime }\wedge d\phi ^{\prime }=r_{0}\sqrt{1+\chi ^{2}}dt\wedge
d\phi $, we get:
\begin{gather}
\mathcal{T}_{0}^{0}= \frac{B}{r_{0}\sqrt{1+\chi ^{2}%
}} -\frac{1+\chi ^{2}}{r_{0}}\ ,  \notag \\
\mathcal{T}_{0}^{2}=-T_{2}^{0}=\frac{\chi \sqrt{1+\chi
^{2}}}{r_{0}}\ ,
\label{rotating_stress} \\
\mathcal{T}_{2}^{2}=\frac{\chi ^{2}}{r_{0}}\ .  \notag
\end{gather}%
The stress-energy tensor (\ref{rotating_stress}) can be written as that of a
perfect fluid with pressure
\begin{equation*}
\mathcal{T}_{b}^{a}=(\rho +p)u^{a}u_{b}+p\,\delta _{b}^{a}\ ,
\end{equation*}%
Writing the fluid velocity vector as $u^{a}=(\sqrt{1+v^{2}},v)$, we get:
\begin{gather*}
(1+v^{2})(\rho +p)-p=\frac{1+\chi ^{2}}{r_{0}}-\frac{B}{r_{0}\sqrt{1+\chi
^{2}}}, \\
v\sqrt{1+v^{2}}(\rho +p)=-\frac{\chi \sqrt{1+\chi ^{2}}}{r_{0}} \\
v^{2}(\rho +p)+p=\frac{\chi ^{2}}{r_{0}}.
\end{gather*}%
So we can express the fluid velocity, the energy density and the pressure in
terms of the integration constants $B$ and $\chi $ as:
\begin{gather}
\rho -p=\frac{1}{r_{0}}\left( 1-B^{\prime }\right) , \\
\rho +p=\frac{1-B^{\prime }+\chi ^{2}}{r_{0}(1+2v^{2})}, \\
1+2v^{2}=\sqrt{1+\frac{4\chi ^{2}(1+\chi ^{2})}{(1-B^{\prime }+\chi
^{2})-4\chi ^{2}(1+\chi ^{2})}}, \\
B^{\prime }\equiv \frac{B}{\sqrt{1+\chi ^{2}}}.
\end{gather}

Therefore as expected, a stationary rotating string must have non-zero
stress. It is worth also pointing out that, in the rotating case the
presence of a tension does not imply $B=0$, so that outside metric is given
by a rotating cone in opposition to the static case \cite{jackiwstring}
where it was found that the outside metric can only correspond to a cylinder.

It can be seen that the rotating string with tension generates inside a flat
space-time with inevitable frame dragging. Writing the geodesic equation for
the outer metric, one can show that a particular solution is $\dot{r}=\dot{%
\phi}=\ddot{t}=0$. Therefore we can use $t$ as an affine parameter so that
\begin{equation*}
\frac{\partial \phi }{\partial t}=\frac{\partial r}{\partial t}=0\qquad
\text{(outside)}.
\end{equation*}%
In the interior, a geodesic observer at constant radius has $\phi ^{\prime
}= $ const. i.e.
\begin{equation*}
\frac{\partial \phi }{\partial t}=-\frac{\chi }{r_{0}}\qquad \text{(inside)}.
\end{equation*}%
So there is a relative angular velocity between inertial observers inside
and outside. For an inertial observer outside, an inertial observer inside
appears accelerated. This is analogous to the case of a rotating shell in
3+1 dimensions, see e. g. \cite{Weinberg}.

\subsection{Incompatibility of a torsion source with a continuous metric}

\label{nogo}

Suppose now that we have a string with spin current, which is a
source of torsion as well as curvature. In this case, the source
corresponds to a continuous distribution of spinning point
particles. In the case of a spinning point particle one has an
exterior geometry given by the spinning cone with a $\delta
$-distribution of torsion at the origin \cite{jackiwlectures}.
Hence, a spinning string with circular symmetry located at some
fixed radius $r_{0}$, possesses the same exterior geometry as the
spinning point particle, but now the distribution of torsion is of
the form $T^{0}=\gamma J\delta (r-r_{0})dr\wedge d\phi $, so that
it reduces to the spinning point particle in the limit of zero
string length. We write the exterior metric in the simple form of
equation (\ref{Rotating-Cone-Metric})
 and we choose the vielbein as
\begin{equation}
e_{+}^{0}=dt-\frac{J}{2}d\phi \ ,\ e_{+}^{1}=d\tilde{r}\ ,\ e_{+}^{2}=B%
\tilde{r}d\phi \ .
\end{equation}%

As it has been discussed above, in this case, the junction
conditions imply that one must have a finite jump in the
tangential components of the metric. However, one may naturally
wonder if there is an alternative possibility of finding a
consistent solution in this situation with a continuous intrinsic
metric. Here we show that this possibility cannot be realized.

Assuming that the metric is continuous at $\Sigma $, in the
presence of above torsion distribution, the spin connection
necessarily acquires a delta function with support on the string
worldsheet. Precisely, the torsion delta function comes only from
a delta function in the contorsion part of the spin connection.
Thus, the spin connection is found to be:
\begin{equation*}
\omega ^{ab}=\mathring{\omega}^{ab}+\kappa ^{ab}
\end{equation*}%
where $\mathring{\omega}^{ab}$ is the Levi-Civita connection of the spinning
cone metric satisfying $de^{a}+\mathring{\omega}^{ab}e_{b}=0$, and the
contorsion $\kappa ^{ab}$ is given by
\begin{gather}
\kappa _{\ 1}^{0}=-\gamma J\delta (r-r_{0})d\phi ,\qquad \kappa _{\ 2}^{0}=%
\frac{\gamma J}{r_{0}}\delta (r-r_{0})dr, \\
\qquad \kappa _{\ 2}^{1}=\frac{\gamma J^{2}}{2r_{0}}\delta (r-r_{0})d\phi -%
\frac{\gamma J}{r_{0}}\delta (r-r_{0})dt.  \notag
\end{gather}%
The curvature 2-form can then be decomposed into a piece $\mathring{\Omega}%
^{ab}=d\mathring{\omega}^{ab}+\mathring{\omega}_{\ c}^{a}\mathring{\omega}%
^{cb}$ which is constructed from the Levi-Civita connection, plus additional
terms including the contorsion as
\begin{equation*}
\Omega ^{ab}=\mathring{\Omega}^{ab}+\mathring{D}\kappa ^{ab}+\kappa _{\
c}^{a}\kappa ^{cb}\ ,
\end{equation*}%
where $\mathring{D}$ stands for the covariant derivative with respect to the
connection $\mathring{\omega}^{ab}$. Therefore, in this way one cannot avoid
the appearance of delta function squared terms in the Riemann tensor. In
particular, since $(\kappa ^{2})_{\ 2}^{0}=\frac{\gamma ^{2}J^{2}}{r_{0}}%
\delta ^{2}(r-r_{0})d\phi \wedge dt$, the component $R_{\
202}^{0}$ contains a delta function squared which is ill-defined
as a distribution. Since we are in 2+1 dimensions this means that
the Einstein tensor possesses a delta function squared, and hence
the field equations imply that the stress-energy tensor is
ill-defined in the distributional sense. This means that the
matter distribution is not physical because it cannot be
normalized. Therefore, one concludes the assumption of having a
continuous tangential metric in this case leads to an unphysical
situation since the matter distribution needed to satisfy the
field equations cannot exist.

Conversely, we conclude that discontinuities in the metric
generically require the presence of torsion concentrated on
$\Sigma$. Indeed, in the absence of torsion, the discontinuity
would imply that the connection contains a delta function, leading
to a delta function squared in the Riemann tensor, as is well
known\cite{geroch,regge_calculus}. We emphasise that torsion may
not be required for special cases, such as travelling
waves\cite{garfinkle}, where the Riemann tensor is linear in the
distributional part of the metric.

This example then shows that, in order to have a physically consistent
solution to this problem, one needs to generalize the standard junction
conditions as in Eqs. (\ref{curvature_junction}), (\ref{torsion_junction}),
which allow for a discontinuity in the tangential metric. This is discussed
in the next subsection.

\subsection{The spinning closed string}

\label{spinning_section}

The metric for the exterior region $M_{+}$ of the spinning string
corresponds to the rotating cone as in Eq.
(\ref{Rotating-Cone-Metric}) but here we use coordinates so that
the spatial section is conformally flat:
\begin{equation}
ds_{+}^{2}=-\left( dt-\frac{J}{2}d\phi \right) ^{2}+
\left(\frac{r}{r_0}\right)^{\! -2(1-B)} \left( dr^{2}+ r^{2}d\phi
^{2}\right)\ ,
\end{equation}%
and we choose the vielbein as
\begin{equation}
e_{+}^{0}=dt-\frac{J}{2}d\phi \ ,\ e_{+}^{1}=
\left(\frac{r}{r_0}\right)^{\! B-1} dr\ , \ e_{+}^{2}=
\left(\frac{r}{r_0}\right)^{\! B-1}r d\phi \ .
\end{equation}%
 For the interior region $M_{-}$, the metric is
flat, and the vielbein is chosen to be
\begin{equation}
e_{-}^{0}=dt^{\prime }\ ,\ e_{-}^{1}=dr\ ,\ e_{-}^{2}=rd\phi ^{\prime }\ .
\end{equation}%
Hence we have a discontinuity in $e^{0}$ across the worldsheet
$\Sigma $. Let us now find the stress-energy $(\mathcal{T}_{\Sigma
})_{a}^{\mu }$ and spin current $(\mathcal{S}_{\Sigma })_{ab}^{\mu
}$ through the junction conditions (\ref{curvature_junction}) and
(\ref{torsion_junction}), respectively.

The position of the string is defined to be $r=r_{0}$. We have
constructed the vielbein so that the normal
$e^1$ is continuous across $\Sigma$.  The junction condition (\ref%
{volume_condition}) relating the induced volume on $\Sigma $
measured from both regions is satisfied choosing $t=t^{\prime }$,
$\phi =\phi ^{\prime }$. So the vielbeins on $\Sigma $ induced by $%
M_{+}$ and $M_{-}$ are of the form
\begin{equation}
\theta _{+}^{0}=dt-\frac{J}{2}d\phi \ ,\ \theta _{+}^{2}=r_{0}d\phi \ ,
\notag
\end{equation}%
and
\begin{equation}
\theta _{-}^{0}=dt\ ,\ \theta _{-}^{2}=r_{0}d\phi \ .  \notag
\end{equation}%
respectively. Using either induced metric, the induced volume element is the
same, i. e.,
\begin{equation*}
\text{Vol}_{\Sigma _{+}}=\text{Vol}_{\Sigma _{-}}=\text{Vol}_{\Sigma
}=-r_{0}dt\wedge d\phi \ .
\end{equation*}%
Note also that the conditions (\ref{extra1}) and (\ref{extra2})
are satisfied.

Since the torsion is zero outside of the string, the only nonvanishing
component of the spin connection are
\begin{gather}
\omega _{+}^{12}=-Bd\phi \ , \\
\omega _{-}^{12}=-d\phi \ ,
\end{gather}%
and since the corresponding curvatures outside of the string
vanish, the vacuum field equations are satisfied.

As required by the junction conditions (\ref{curvature_junction})
and (\ref{torsion_junction}), in order to find the stress-energy and spin current on $%
\Sigma $ we need the discontinuities of the vielbein and connection pulled
back to the tangent space of $\Sigma $, i.e. $i^{\ast }\Delta e^{a}=\Delta
\theta ^{a}$ and $i^{\ast }\Delta \omega ^{ab}$, whose only non-vanishing
components are
\begin{gather}
i^{\ast }\Delta e^{0}=-\frac{J}{2}d\phi \ , \\
i^{\ast }\Delta \omega ^{12}=(1-B)d\phi \ .
\end{gather}%
Substituting these into the junction conditions
(\ref{curvature_junction}) and (\ref{torsion_junction}), we find
that the only non-vanishing component of the stress-energy tensor
is
\begin{equation}
(\mathcal{T}_{\Sigma })_{0}^{0}=-\frac{(1-B)}{r_{0}}\ ,
\label{spinnig-stress-energy}
\end{equation}%
and analogously, for the spin current one obtains
\begin{equation}
(\mathcal{S}_{\Sigma })_{12}^{0}=-(\mathcal{S}_{\Sigma })_{21}^{0}=\frac{J}{%
2r_{0}}.  \label{spinnig-current}
\end{equation}%
Therefore, the solution describes a string made of spinning dust.

In sum, the metric for the spinning string in the whole spacetime can be
written as%
\begin{equation}
ds^{2}=- \left( dt-\frac{J}{2}\theta (r-r_{0})d\phi \right)
^{2}+\left(1+\left\{\left(\frac{r}{r_0}\right)^{-2(1-B)}-1\right\}\theta
(r-r_{0}) \right)(dr^{2}+r^{2}d\phi ^{2})\ ,
\label{spinning-string-metric}
\end{equation}%
where the only nonvanishing component of the torsion generated by the spin
current turns out to be%
\begin{equation*}
T^{0}=-\frac{J}{2}\delta (r-r_{0})dr\wedge d\phi \ .
\end{equation*}%
This can be explicitly checked writing the vielbein as
\begin{eqnarray*}
e^{0} &=&dt-\frac{J}{2}\theta (r-r_{0})d\phi \ , \\
e^{1}
&=&\left(1+\left\{\left(\frac{r}{r_0}\right)^{-(1-B)}-1\right\}\theta
(r-r_{0}) \right) dr\ , \\
e^{2}
&=&\left(1+\left\{\left(\frac{r}{r_0}\right)^{-(1-B)}-1\right\}\theta
(r-r_{0}) \right)rd\phi \ ,
\end{eqnarray*}%
and using the fact that the only non zero components of the spin connection
are the same as in Eq. (\ref{omega12}).

Some interesting remarks can be made comparing the spinning dust
string with the rotating string with tension discussed in
\ref{rotating_section}. Since the metric of the exterior region is
the same in both cases (up to a co-ordinate rescaling of $r$ by a
constant factor), the total mass and angular momentum coincide.
However, the source of the angular momentum has a different nature
in both cases. For the rotating string with tension the angular
momentum $J$ is a consequence of the fact that fluid is stationary
with a given angular velocity, while for the spinning string, the
non-zero angular momentum of the exterior solution is caused by
the torsion produced by the distribution of static spinning dust
particles. Furthermore, as opposed to the rotating string with
tension, where the inner flat space-time has an inevitable frame
dragging, for the spinning string the inertial frames in the
interior are not dragged. It is then worth pointing out that in
this sense, the dragging effect can be \textquotedblleft shielded"
by considering spinning instead of rotating sources.

Let us digress for a moment to discuss the point particle
\cite{jackiwlectures}. In this case the metric does not
distinguish between the spinning and rotating sources. In both
cases we have a delta in the energy density. The source of angular
momentum can either be a delta function in the torsion or a
$\mathcal{T}_{0\phi}$ is the derivative of a delta function (point
particle limit for a rotating object). If the source  acquires a
bit of length the metric distinguishes between the two situations.

Regarding the spinning string as a limiting case of some finite
distribution of matter with aligned spin, makes the metric change
very rapidly across the thickness of the string. In this sense,
one concludes that the spin has a more dramatic effect on the
geometry than does the mass: The mass causes the first derivative
of the connection to blow up whereas the spin causes the first
derivative of the metric to blow up.

As discussed in the next section, these effects can be seen to occur also in
four and higher dimensions.

\section{Four and higher dimensional case}\label{higher_chapter}

The Lagrangian for General Relativity in first order formalism in arbitrary
dimension $D$ is%
\begin{equation*}
\mathcal{L}=\frac{1}{(D-2)!}\Omega ^{a_{1}a_{2}}e^{a_{3}}\cdot \cdot \cdot
e^{a_{D}}\epsilon _{a_{1}a_{2}a_{3}\cdot \cdot \cdot a_{D}}\ .
\end{equation*}%
Hereafter wedge product between forms is understood. The variation with
respect to the vielbein gives the Einstein field equations%
\begin{equation}
\Omega ^{bc}\,dx^{\mu }\, \text{Vol}_{abc}=-2\mathcal{T}_{a}^{\mu
}\text{Vol\ ,}  \label{higherEinstein}
\end{equation}%
and the variation with respect to the spin connection allows to fix the
torsion in terms of the spin current as
\begin{equation}
dx^{\mu }T^{c}\text{Vol}_{abc}=\mathcal{S}_{ab}^{\mu }\text{Vol\ ,}
\label{higherSpin}
\end{equation}%
where
\begin{equation}
\text{Vol}_{abc}:=\frac{1}{(D-3)!}\overbrace{e^{a_{4}}\cdots e^{a_{D}}}^{D-3%
\text{ times}}\epsilon _{abca_{4}\cdots a_{D}}  \label{Vol-abc}
\end{equation}%
is a volume element of a $d-3$-dimensional surface orthogonal to $%
e^{a}\wedge e^{b}\wedge e^{c}$. It is useful to define also%
\begin{equation*}
\text{Vol}_{a_{1}\cdot \cdot \cdot a_{n}}:=\frac{1}{(D-n)!}\overbrace{%
e^{a_{n+1}}\cdots e^{a_{D}}}^{D-n\text{ times}}\epsilon _{a_{1}\cdots a_{D}}
\end{equation*}%
\newline

\subsection{Junction conditions in higher dimensions}

We derive the junction conditions in $D>3$ dimensions. The geometry is
assumed to be smooth on each side with a possible discontinuity on the
domain wall $\Sigma $, thus we have smooth fields $e_{+}^{a}$, $\omega
_{+}^{ab}$ on one side and smooth fields $e_{-}^{a}$, $\omega _{-}^{ab}$ on
the other side. The geometry of the entire manifold is given then by some
distributional vielbein and spin connection which coincide with the smooth
functions $e_{\pm }^{a}$ and $\omega _{\pm }^{ab}$ outside the surface $%
\Sigma $.

Furthermore, we shall assume that there is a well defined normal
vector on the hypersurface.

The distribution $e^{a}$ can be defined by a sequence of smooth vielbeins $%
e_{\alpha }^{a}$ which interpolate between the discontinuous
values within some neighbourhood $O_{\alpha }$ of the hypersurface
$\Sigma $. The width of region $O_{\alpha }$ is of order $1/\alpha
$ so that the distribution $e^{a}$ is obtained in the limit
$\alpha \rightarrow \infty $. The distributional spin connection
is defined in the same way. We assume that the fields $ e_{\pm
}^{a}$ and $\omega _{\pm }^{ab}$ can be continued smoothly across
region $O_{\alpha }$.

It is useful to consider the
classes of vielbeins and spin connections defined as%
\begin{eqnarray}
\lbrack e^{a}]_{\beta } &=&e_{-}^{a}+\beta \Delta e^{a}\ ,  \notag \\
\lbrack {\omega }^{ab}]_{\gamma } &=&\omega _{-}^{ab}+\gamma
\Delta \omega ^{ab}\ ,  \label{e-class}
\end{eqnarray}%
which by definition are smooth one-forms, for some arbitrary
constants $\beta $,$\gamma $. Then, without any loss of
generality, we can decompose $e_{\alpha }^{a}$ and ${\omega
_{\alpha }^{ab}}$ as follows:
\begin{gather}
{e_{\alpha }}^{a}=[e^{a}]_{\beta }+y_{\alpha ,\beta }\Delta
e^{a}+z_{\alpha
}^{a}\ , \label{expand_e_alpha}\\
{\omega _{\alpha }}^{ab}=[{\omega }^{ab}]_{\gamma }+g_{\alpha
,\gamma }\Delta \omega ^{ab}+h_{\alpha }^{ab}\
,\label{expand_omega_alpha}
\end{gather}%
where $z_{\alpha }^{a}$ and $h_{\alpha }^{ab}$ are smooth one
forms which vanish on $\partial O_{\alpha }$ and outside of
${O_{\alpha }}$, and the functions $y_{\alpha ,\beta }$ and
$g_{\alpha ,\gamma }$ tend to the distribution $\theta (\Sigma
)-\beta $, and $\theta (\Sigma )-\gamma $, respectively.

\subsubsection{Constraints for well defined sources}
Let us look at the right hand side of the Einstein equation
(\ref{higherEinstein}). We wish to integrate over the
infinitesimal region and take the limit
\begin{gather*}
 \lim_{\alpha \to \infty}\int_{O_\alpha} ({\cal T}_\alpha)^\mu_a \text{Vol}_\alpha
\end{gather*}
where $\text{Vol}_\alpha$ is the volume element constructed with
$e_\alpha$.

To each $e_\alpha$ and $\omega_\alpha$ in the sequence, there is a
corresponding stress-energy tensor $({\cal T}_\alpha)^\mu_a$. In
the limit $\alpha \to \infty$ we require that $({\cal
T}_\alpha)^\mu_a$ becomes a delta function distribution.

Expanding Vol$_\alpha$ according to equation
(\ref{expand_e_alpha}) we obtain
\begin{gather*}
 \frac{1}{(D-1)!}\int N_z dz \int ({\cal T}_\alpha)^\mu_a ([e]_\beta +
 y_{\alpha,\beta}\Delta e + z_\alpha)^{a_2} \cdots
 ([e]_\beta +
 y_{\alpha,\beta}\Delta e + z_\alpha)^{a_D} \epsilon_{b a_2 \cdots
 a_D} N^b
\end{gather*}
Above we have split the vielbein in terms of a normal one-form
$e^1 = N_z dz$ where $z$ is a co-ordinate normal to the
hypersurface, and $e^2, \dots e^D$, which are tangential. Note
that $N^b \equiv \delta^b_1$.

We require the integral to be independent of the limiting process
and converge to
\begin{gather}
 \int_{\Sigma} ({\cal T}_\Sigma)^\mu_a \text{Vol}_\Sigma,
\end{gather}
with \begin{gather}
 \text{Vol}_\Sigma := \frac{1}{(D-1)!}e^{a_2}\cdots
e^{a_D}\epsilon_{b a_2 \cdots a_D}N^b
 .
\end{gather}
Since the integral of the product $({\cal T}_\alpha)^\mu_a$ times
$y_{\alpha, \beta}$ to some power depends on the limiting process,
we need to require that such terms vanish identically. We assume
that $z_\alpha$ tends to zero smoothly enough so that it gives
vanishing contribution to the integral in the limit. Consequently,
$([e]+y\Delta e)^{D-1}$ must be independent of the function $y$.
This can be stated in three equivalent ways:\\ {\bf
i)}\begin{gather}
 \frac{d}{d\gamma} \text{Vol}_{\gamma,a} N^a = 0
\end{gather}
where $\text{Vol}_{\gamma,a}$ is $\text{Vol}_{a}$ constructed with
$[e]_\gamma^a$;
\\{\bf ii)}\begin{gather}\label{delta_volumes}
 \Delta e^{a_1}\cdots \Delta e^{a_p} \Delta\text{Vol}_{ba_1 \cdots a_p}N^b =0,
 \qquad
 \forall\ 0\leq p\leq D-2;
\end{gather}
{\bf iii)}
\begin{gather}
 e_+^{a_2}  \cdots e_+^{a_D}\epsilon_{b a_2 \cdots a_D}N^b
 \ =\ e_+^{a_2}\cdots e_+^{a_{D-1}} e_-^{a_D}\epsilon_{ba_1 \cdots a_D}N^b
 \ = \cdots\nonumber
 \\ \cdots\ =\ e_-^{a_2}\cdots e_-^{a_D}\epsilon_{ba_1 \cdots a_D}N^b.
 \end{gather}
This last expression means that the intrinsic volume element
$\text{Vol}_\Sigma$ must be invariant under the operation of
swapping $e_+$ with $e_-$ anywhere in the product. In particular,
this means that $\text{Vol}_{\Sigma_+}= \text{Vol}_{\Sigma_-}$.

Notice that in the special case of 2+1 dimensions, we have two
conditions: $\text{Vol}_{\Sigma_+} = \text{Vol}_{\Sigma_-}$ and
$i^*\Delta e^{[a} \Delta e^{b]} =0$, as can be seen from equation
(\ref{delta_volumes}).

Looking at the equation of the spin current, we obtain the same
conditions.

\subsubsection{Junction conditions from the field equations}

Now let us look at the left hand side of the torsion equation
given by (\ref{higherSpin}).
Integrating this over the region $O_{\alpha }$ one obtains%
\begin{equation*}
 \int_{O_{\alpha }}dx^{\mu }T_{\alpha }^{c}\text{Vol}_{\alpha
 ,abc}=\int_{\partial O_{\alpha }}dx^{\mu }\text{Vol}_{\alpha
 ,ab}+(\cdots )\,
\end{equation*}%
In the limit that the region $O_{\alpha }$ shrinks to zero thickness, only
the terms involving $de_{\alpha }^{a}$ contributes to this integral, as we
suppose the discontinuities to be finite. Thus, the dots $(\cdots )$
represents those terms which vanish in the limit. Here Vol$_{\alpha ,abc}$
stands for the volume in Eq. (\ref{Vol-abc}) constructed with $e_{\alpha
}^{a}$.

Therefore, recalling Eq. (\ref{DistributionSpin}), which is valid for any
dimension, the junction condition for the spin current is
\begin{equation}
i^{\ast }\left( dx^{\mu }\Delta \text{Vol}_{ab}\right) =(\mathcal{S}_{\Sigma
})_{ab}^{\mu }\text{Vol}_{\Sigma }\ ,  \label{junction-torsion-D}
\end{equation}%
where $\Delta $Vol$_{ab}:=$Vol$_{+,ab}-$Vol$_{-,ab}\ $. In the
three-dimensional case this reduces to Eq. (\ref{torsion_junction}).

Now, let us look at the Einstein equation (\ref{higherEinstein}). We must
integrate
\begin{equation}
I=\int_{O_{\alpha }}\Omega _{\alpha }^{bc}\,\text{Vol}_{\alpha ,abc}\,f^{a}
\label{Integral-Einstein}
\end{equation}%
where we have introduced $f^{a}=f_{\mu }^{a}dx^{\mu }$ as some arbitrary
smooth test one-form. In the limit that the region $O_{\alpha }$ shrinks to
zero thickness, only terms involving $d\omega _{\alpha }$ or $de_{\alpha }$
will contribute to (\ref{Integral-Einstein}). Thus the integral (\ref%
{Integral-Einstein}) reads
\begin{eqnarray}
I &=&\int_{O_{\alpha }}d\omega _{\alpha }^{bc}\text{Vol}_{\alpha ,abc}\
f^{a}+(\cdots )\   \notag \\
&=&\int_{\partial O_{\alpha }}\omega _{\alpha }^{bc}\text{Vol}_{\alpha
,abc}\ f^{a}+\int_{O_{\alpha }}\omega _{\alpha }^{bc}d\text{Vol}_{\alpha
,abc}\ f^{a}+(\cdots )\ ,  \label{I2} \\
&=&B_{1}+V+(\cdots )\ ,  \notag
\end{eqnarray}%
The boundary term $B_{1}$ in Eq. (\ref{I2}) in the limit turns out to be%
\begin{equation}
B_{1}=\int_{\Sigma }\left( \omega _{+}^{bc}\text{Vol}_{+,abc}-\omega
_{-}^{bc}\text{Vol}_{-,abc}\right) \ f^{a}\ ,  \label{Boundary-D}
\end{equation}%
which is expected to contribute to the junction conditions. The remaining
volume integral $V$ must be handled with care, since it will give a non-zero
contribution to the junction conditions.

Now let us expand $\omega_\alpha$ and $e_\alpha$ according to
equations (\ref{expand_e_alpha}) and (\ref{expand_omega_alpha}).
 Thus, the remaining volume integral reads%
\begin{eqnarray*}
V &=&\int_{O_{\alpha }}\left( [{\omega }^{bc}]_{\gamma }+h_{\alpha
}^{bc}\right) d\text{Vol}_{\alpha ,abc}\ f^{a} \\
&&-\left( g_{\alpha ,\gamma }dy_{\alpha ,\beta }\right) \ \Delta \omega
^{bc}\Delta e^{d}\ \text{Vol}_{\alpha ,abcd}\ \ f^{a}+(\cdots )\ ,
\end{eqnarray*}%
and since $h_{\alpha }^{bc}$ vanishes at $\partial O_{\alpha }$, integrating
by parts the first term we obtain%
\begin{equation*}
V=B_{2}-\left( g_{\alpha ,\gamma }dy_{\alpha ,\beta }\right) \ \Delta \omega
^{bc}\Delta e^{d}\ \text{Vol}_{\alpha ,abcd}\ f^{a}+(\cdots )\ ,
\end{equation*}%
where the boundary term $B_{2}$ in the limit is given by

\begin{equation}
B_{2}:=-\int_{\Sigma }[{\omega }^{bc}]_{\gamma }\Delta \text{Vol}_{abc}\
f^{a}\ .  \label{B2}
\end{equation}
The only remaining part of the volume integral $V$ which does not vanish in
the limit is:%
\begin{equation*}
-\int_{O_{\alpha }}\left\{ g_{\alpha ,\gamma }dy_{\alpha ,\beta }\right\} \
\Delta \omega ^{\lbrack bc}\Delta e^{d]}\ \text{Vol}_{\alpha ,abcd}\ \
f^{a}\ ,
\end{equation*}%
where the object in the curly brackets depends on the limiting
process. Therefore, in order to have a result which is independent
of the limiting process, one needs to impose the following
condition:
\begin{equation}
i^{\ast }(\Delta \omega ^{\lbrack ab}\wedge \Delta e^{c]})=0\ .
\label{theConstraint}
\end{equation}%
Note that only the pull-back $i^{\ast }$ onto the surface $\Sigma $ appears
because in the limit $\alpha \rightarrow \infty $ only the normal
derivatives of $y$ blows up.

Since $\Delta $Vol$_{abc}$ can be expanded as\footnote{%
The explicit form of $X_{abcd}$ is%
\begin{equation*}
X_{abcd}=\frac{1}{(D-3)!}\epsilon _{abcda_{5}\cdot \cdot \cdot
a_{D}}\sum_{p=0}^{D-4}e_{+}^{a_{5}}\cdot \cdot \cdot
e_{+}^{a_{5+p}}e_{-}^{a_{6+p}}\cdot \cdot \cdot e_{-}^{a_{D}}\ .
\end{equation*}%
}%
\begin{equation*}
\Delta \text{Vol}_{abc}=\Delta e^{d}X_{abcd}\ ,
\end{equation*}%
the boundary term $B_{2}$ reads%
\begin{equation*}
B_{2}:=-\int_{\Sigma }\left( {\omega }_{-}^{bc}\Delta \text{Vol}%
_{abc}+\gamma \Delta \omega ^{\lbrack bc}\Delta e^{d]}\ X_{abcd}\right) \
f^{a}\ ,
\end{equation*}%
so that the second term vanishes by virtue of the condition (\ref%
{theConstraint}), and therefore the original integral (\ref%
{Integral-Einstein}) reads%
\begin{eqnarray*}
I &=&B_{1}+B_{2} \\
&=&\int_{\Sigma }\Delta {\omega }^{bc}\text{Vol}_{+,abc}\ f^{a}\ ,
\end{eqnarray*}%
Note that because of the condition (\ref{theConstraint}) one can see that
\begin{equation*}
\Delta \omega ^{\lbrack ab}e_{+}^{c]}=\Delta \omega ^{\lbrack
ab}[e^{c]}]_{\gamma }\ ,
\end{equation*}%
and hence, the value of $I$ is the same if instead of using Vol$_{+,abc}$
computed with $e_{+}^{a}$ , one uses Vol$_{\gamma ,abc}$ which is computed
with any representative of the class $[e^{a}]_{\gamma }$ defined as in Eq. (%
\ref{e-class}). Therefore the last junction condition is%
\begin{equation}
i^{\ast }\left( \Delta {\omega }^{bc}\,dx^{\mu
}\,\text{Vol}_{\gamma ,abc} \right) =-2(\mathcal{T}_{\Sigma
})_{a}^{\mu }\text{Vol}_{\Sigma }\ . \label{curvature-junction-D}
\end{equation}

\subsubsection{Summary}

Note that the junction condition can be written in purely
anholonomic indices even without an induced metric.
Because of the
constraint (\ref{theConstraint}), we can contract equation
(\ref{curvature-junction-D}) with $e_{\mu }^{a}$
 which gives
\begin{equation} i^{\ast }\left( \Delta {\omega
}^{cd}\,e^b\,\text{Vol}_{acd} \right) =-2(\mathcal{T}_{\Sigma
})_{a}^{b }\text{Vol}_{\Sigma }\ .\label{junction_summary_1}
\end{equation}
where the same result is obtained by contracting with any
$e^a_\mu$ in the class (\ref{e-class}).

Analogously, the junction condition for the spin connection can be
written with purely anholonomic indices because of condition
$i^{\ast }\Delta e^{a}\Delta \!\text{Vol}_{ab}=0$ from equation
(\ref{delta_volumes}). It is:
\begin{equation}
i^{\ast }e^d\,\Delta \!\text{Vol}_{ab}=(\mathcal{S}_{\Sigma
})_{ab}^{^d }\label{junction_summary_2}
\end{equation}

In sum we have a hypersurface $\Sigma$ on which the tangential
vielbeins and the spin connection have a bounded discontinuity and
the normal is assumed to be continuous across $\Sigma$. The set of
conditions for General Relativity in the presence of spinning
sources is given by (\ref{junction_summary_1}) and
(\ref{junction_summary_2}) with the purely geometrical condition
\begin{gather}
 \Delta e^{a_1}\cdots \Delta e^{a_p} \Delta\text{Vol}_{ba_1 \cdots a_p}N^b  =0,
 \qquad
 \forall\ 0\leq p\leq D-2, \label{constraint_summary_1}
\end{gather}
ensuring that the volume element on $\Sigma$ does not depend on
the induced metric, together with the constraint
\begin{gather}
 \Delta \omega^{[ab} \Delta e^{c]} =
 0\label{constraint_summary_2}.
\end{gather}

From the internal consistency of the junction conditions
(\ref{junction_summary_1}), (\ref{junction_summary_2}) and
(\ref{constraint_summary_2}) we get:
\begin{gather}
\Delta e_{\mu }^{a}(\mathcal{T}_{\Sigma })_{a}^{\mu }=0,
\\
\Delta \omega _{\mu }^{ab}(\mathcal{S}_{\Sigma })_{ab}^{\mu }=0.
\label{delta_omega_T}
\end{gather}
But we note that $\Delta \omega^{ab}_\mu$ is related to $({\cal
T}_\Sigma)^\mu_a$ by:
\begin{gather}
 ({\cal T}_\Sigma)^i_j = \Delta \omega^{1 i}_{\ j} - \Delta
 \omega^{1k}_{\ k} \delta^i_j,
 \\
 ({\cal T}_\Sigma)^i_1 = \Delta \omega^{ik}_{\ k}.
\end{gather}
where we use the frames adapted to the hypersurface so that $e^1$
is the normal vector and $e^i$, $i =2,\dots, D$ are tangential.
Putting the above equations in (\ref{delta_omega_T}), we obtain
the following invariant equation
\begin{gather}
\label{the_beautiful_equation}
 N^a \left(({\cal T}_\Sigma)^b_c - \frac{1}{D-2}\ \delta^b_c {\cal
 T}_\Sigma \right) (S_\Sigma)^c_{ab} = 0.
\end{gather}
Remarkably, this imposes a constraint between the direction of
flow of energy and the orientation of the spin. Very roughly
speaking, this is a kind of polarisation condition.

In order to evaluate the junction conditions, we need to use
co-ordinate patches which are smooth across $\Sigma$.

%This is somewhat similar to the status of the junction conditions
%for a continuous metric before the paper of Israel \cite{israel}.

%This suggests that there exists a well defined action principle describing
%the discontinuous geometry. $\delta S=\delta e_{\mu }^{a}(\mathcal{T}%
%_{\Sigma })_{a}^{\mu }+\delta \omega _{\mu
%}^{ab}(\mathcal{S}_{\Sigma })_{ab}^{\mu }=0$ where the $\delta e$
%and $\delta \omega $ are unambiguous******in the sense that the
%variation can be done for any guy in the class**

The junction conditions for Einstein-Cartan and Lovelock-Cartan
gravity and the action principle will be analysed in detail in a
future publication\cite{future}.
\newline

\subsection{Rotating vs. spinning cylinders in four dimensions}

As an application, we consider a $3+1$-dimensional example. Let us
consider the case of a straight spinning cylinder, which can be
obtained as the lifting of the spinning string of Section
\ref{spinning_section} by introducing an extra $z$ direction, with
$\partial_z$ a killing vector. The exterior metric is
\begin{equation*}
ds_{+}^{2}=-\left( dt-\frac{J}{2}d\phi \right) ^{2}+
\left(\frac{r}{r_0}\right)^{\! -2(1-B)} \left( dr^{2}+ r^{2}d\phi
^{2}\right) +dz^2
\end{equation*}%
and we choose the vielbein as
\begin{equation*}
e_{+}^{0}=dt-\frac{J}{2}d\phi \ ,\ e_{+}^{1}=
\left(\frac{r}{r_0}\right)^{\! B-1} dr\ , \ e_{+}^{2}=
\left(\frac{r}{r_0}\right)^{\! B-1}r d\phi\ ,\ e_+^3 = dz .
\end{equation*}%
 For the interior region $M_{-}$, the metric is
flat, and the vielbein is chosen to be
\begin{equation*}
e_{-}^{0}=dt\ ,\ e_{-}^{1}=dr\ ,\ e_{-}^{2}=rd\phi\ ,\ e^3_- = dz
.
\end{equation*}%

There is a single discontinuous component of the spin connection
and one discontinuous component of the $i^*\text{Vol}_{ab}$,
\begin{gather*}
i^*\Delta \text{Vol}_{12} = -i^*\Delta \text{Vol}_{21} = -\frac{J}{2}d\phi\wedge dz \ , \\
i^{\ast }\Delta \omega ^{12}=(1-B)d\phi \ .
\end{gather*}
and the other components vanish. It is easy to verify that the
conditions (\ref{constraint_summary_1}) and
(\ref{constraint_summary_2}) are satisfied. Therefore this
solution is compatible with the junction conditions. The
computation of the stress tensor and the spin current is then
straightforward. Using the junction conditions
(\ref{junction_summary_1}) and (\ref{junction_summary_2}) and
noting that Vol$_\Sigma = -r_0 dt\wedge d\phi \wedge dz$, we get
for the stress tensor:
\begin{gather*}
 ({\cal T}_\Sigma)^0_0 =
 ({\cal T}_\Sigma)^3_3 = -\frac{1-B}{r_0},
\end{gather*}
This is the same result as in $2+1$ but there is a pressure along
the length of the cylinder as naturally expected. The spin current
is given by:
\begin{gather*}
 (S_\Sigma)^0_{12}= -(S_\Sigma)^0_{21} = \frac{J}{2r_0}.
\end{gather*}

To summarise, the metric for the spinning cylinder in the whole
spacetime can be written as%
\begin{equation}
ds^{2}= ds^2_{(3)} +dz^2\ , \label{spinning-string-metric_4d}
\end{equation}%
where $ds^2_{(3)}$ is given by equation
(\ref{spinning-string-metric}). The only nonvanishing component of
the torsion generated by
the spin current turns out to be%
\begin{equation*}
T^{0}=-\frac{J}{2}\delta (r-r_{0})dr\wedge d\phi \ .
\end{equation*}%

Analogously, the rotating cylinder can be solved as a
straightforward extension of the rotating string in 2+1 dimensions
found in Section \ref{rotating_section}. The stress tensor is the
same as in equation (\ref{rotating_stress}) but with a pressure
$({\cal T}_\Sigma)^3_3$ along the length of the cylinder
satisfying $({\cal T}_\Sigma)^3_3 = ({\cal T}_\Sigma)^0_0$. The
result is in agreement with that found in reference
\cite{flowerpot}. Since these solutions are essentially the same
as in the 2+1-dimensional case, the same effect regarding the
shielding of frame dragging occurs. The generalisation of these
results to higher dimensional rotating and spinning domain walls
with a worldsheet geometry given by $S^1 \times \mathbb{R}^{D-2}$
in $D$-dimensions is straightforward.

\section{Summary and conclusions}

The junction conditions for General Relativity in the presence of
domain walls with intrinsic spin were derived in three and higher
dimensions.

We considered the domain wall as the thin shell limit of some
finite distribution of matter with aligned spin. We required the
independence of the junction conditions on the limiting process
i.e. that a sufficiently thin shell can be approximated by a shell
of strictly zero thickness.

We have seen that the metric must change very rapidly across the
thickness of the domain wall. In this sense, the spin has a more
dramatic effect on the geometry than does the mass: The mass
causes the first derivative of the connection to blow up whereas
the spin causes the first derivative of the metric to blow up.

In general then, when the torsion is localized on the domain wall,
in the zero thickness limit it is necessary to relax the
continuity of the tangential components of the vielbein.

It was shown that a stress tensor and a spin current can be
defined just by requiring the existence of a well defined volume
element which is independent of an induced metric, so as to allow
for generic torsion sources. In fact it was found that the spin
current is proportional to the jump in the vielbein (see equation
\ref{junction_summary_2}) and the stress-energy tensor is
proportional to the jump in the spin connection (equation
\ref{junction_summary_1}).

The consistency of the junction conditions  implies a non-trivial
constraint involving the product of the spin current and stress
tensor, equation (\ref{the_beautiful_equation}). This is a
constraint between the direction of flow of energy and the
orientation of the spin. Very roughly speaking, this is a kind of
polarisation condition.

As an application, we derive the circularly symmetric solutions
for both the rotating string with tension and the spinning dust
string in three dimensions. The rotating string with tension
generates a rotating truncated cone outside and a flat space-time
with inevitable frame dragging inside. In the case of a string
made of spinning dust, in opposition to the previous case no frame
dragging is present inside, so that in this sense, the dragging
effect can be \textquotedblleft shielded" by considering spinning
instead of rotating sources. Applying the junction conditions for
 General Relativity in four dimensions, we found that the previously described
string solutions can be lifted to the rotating and spinning
cylinder with pressure along its length. The generalisation to
higher dimensions is straightforward.

\paragraph{Acknowledgements}

We would like to thank E. Gravanis for discussions on junction
conditions. We also thank Eloy Ay\'{o}n-Beato, Julio Oliva, Jean
Krisch, Pablo Mora and Jorge Zanelli for helpful comments. We
thank F. W. Hehl for bringing to our attention Ref.
\cite{Arkuszewski} where junction conditions for a boundary layer
between two spin sources in the bulk were discussed. This work was
partially funded by FONDECYT grants 1040921, 1051056, 1061291 and
3060016. The generous support to CECS by Empresas CMPC is also
acknowledged. CECS is a Millennium Science Institute and is funded
in part by grants from Fundaci\'{o}n Andes and the Tinker
Foundation.


\thebibliography{99}


\bibitem{israel}
W. Israel, Singular hypersurfaces and thin shells in general
relativity; Nuovo Cim. B44:1 (1966); Erratum Ibid {\bf B48}: 463,
(1967).

\bibitem{marolf}
D. Marolf and S. Yaida, Energy Conditions and Junction Conditions;
Phys. Rev. {\bf D72}, 044016 (2005), gr-qc/0505048.

\bibitem{jackiwstring}
S. Deser, R. Jackiw, String sources in 2+1-dimensional gravity;
Ann. Phys. 192:352 (1989).

\bibitem{Clement}
G. Clement, Annals Phys. 201: 241-257 (1990).


\bibitem{jackiwlectures}
R. Jackiw, Five lectures on planar gravity; Cocoyoc1990
proceedings, relativity and gravitation: classical and quantum
p.74-97, World Scientific (1991); R. Jackiw, Lower dimensional
gravity; Nucl. Phys. B252 343-356 (1985).

\bibitem{cosmicstring1}
A. Vilenkin, Gravitational fields of vacuum domain walls and
strings;
 Phys. rev D23, 852 (1981).

\bibitem{cosmicstring2}
W.A. Hiscock, Exact gravitational field of a string; Phys. Rev.
D31 3288-3290 (1985).

\bibitem{deserjackiw}
S. Deser, R. Jackiw, G. 't Hooft, Three dimensional Gravity:
dynamics of flat space; Ann. Phys. 152, 220-235 (1984).

\bibitem{Weinberg}
S. Weinberg, Gravitation and Cosmology: Principles and
applications of the theory of relativity; John Wiley \& Sons Inc.
New York (1972).

\bibitem{geroch}
R. Geroch, J.H. Traschen, Strings and other distributional sources
in general relativity; Phys. Rev. D36: 1017 (1987).

\bibitem{regge_calculus}
C. Wainwright, Ruth M. Williams, Area Regge calculus and
discontinuous metrics; Class. Quant. Grav. {\bf 21}, 4865-4880
(2004), gr-qc/0405031.

\bibitem{garfinkle}
D.~Garfinkle, ``Metrics with distributional curvature,'' Class.\
Quant.\ Grav.\  {\bf 16}, 4101 (1999) [arXiv:gr-qc/9906053].
%%CITATION = GR-QC 9906053;%%

\bibitem{future}
E. Gravanis and S. Willison, In preparation.

\bibitem{flowerpot}
B. Jensen, H. H. Soleng, General relativistic model of a spinning
cosmic string; Phys. Rev. D45 3528-3533 (1992).

\bibitem{Arkuszewski}
W. Arkuszewski, W. Kopczynski and V. N. Ponomarev, Commun. Math.
Phys. {\bf 45}: 183-190 (1975).
\end{document}